# Observation of Dirac Cone Warping and Chirality Effects in Silicene

Baojie Feng[1], Hui Li[1], Cheng-Cheng Liu[1], Tingna Shao[1], Peng Cheng[1], Yugui Yao[2], Sheng Meng[1], Lan Chen[1]* and Kehui Wu[1]*

[1]*Institute of Physics, Chinese Academy of Sciences, Beijing 100190, China*

[2]*School of Physics, Beijing institute of technology, Beijing 100081, China*

*Corresponding Authors: lchen@iphy.ac.cn (L. Chen) & khwu@iphy.ac.cn (K. H. Wu).

**ABSTRACT   We performed low temperature scanning tunneling microscopy (STM) and spectroscopy (STS) studies on the electronic properties of ($\sqrt{3}\times\sqrt{3}$)R30° phase of silicene on Ag(111) surface. We found the existence of Dirac Fermion chirality through the observation of -1.5 and -1.0 power law decay of quasiparticle interference (QPI) patterns. Moreover, in contrast to the trigonal warping of Dirac cone in graphene, we found that the Dirac cone of silicene is hexagonally warped, which is further confirmed by density functional calculations and explained by the unique superstructure of silicene. Our results demonstrate that the ($\sqrt{3}\times\sqrt{3}$)R30° phase is an ideal system to investigate the unique Dirac Fermion properties of silicene.**

Silicene,[1-3] a single sheet of Si atoms arranged in a honeycomb lattice analogous to graphene, has been successfully fabricated on Ag(111) and other surfaces recently.[4-12] The experiments revealed similarities between silicene and graphene: honeycomb atomic structure, linear dispersion of electron band as well as high Fermi velocity ($10^6$ m/s).[6,7] However, there are still debates. For example, recently C. Lin *et al.* claimed that Dirac fermion is absent in the 4×4 silicene (or 3×3 with respect to silicene-1×1) on Ag(111) surface due to significant symmetry breaking.[13] We note that a linear dispersion near Fermi level is far from a complete description of the Dirac fermion state. In Dirac systems such as graphene or three dimensional (3D) topological insulators, existence of quasiparticle chirality is even more important. Quasiparticle chirality induces exotic quantum phenomena such as Klein tunneling,[14,15] Veselago lens,[16] and suppression of backscattering.[17-21] The latter phenomena, not found in any conventional 2D electron systems, can be regarded as fingerprints of Dirac systems with quasiparticle chirality.

On the other hand, the Dirac cone structure in graphene electron band is directly associated with its hexagonal honeycomb lattice symmetry. Free-standing silicene has a low-buckled structure[1-3] where the two inequivalent sublattices A and B in silicene are not coplanar. It reduces the crystal symmetry from $C_{6v}$ to $C_{3v}$. Moreover, recent experiments and theoretical calculations revealed various surface reconstructions, such as 3×3, ($\sqrt{7}\times\sqrt{7}$)R19.1°, ($\sqrt{3}\times\sqrt{3}$)R30° with respect to silicene-1×1.[6-8,22,23] Many of these structures exhibit reduced symmetry. The influence of these structural features on the Dirac cone structure of silicene remains an open question.

Here, using low temperature scanning tunneling microscopy (STM) and spectroscopy (STS), we found evidence of Dirac fermion chirality by the observation of stronger power law decay of quasiparticle interference (QPI) patterns than that in two dimensional electron gas (2DEG). Moreover, we show that the Dirac cone of silicene is not circular, but significantly warped. In contrast to the trigonal warping of Dirac cone in graphene, the warping in silicene is hexagonal. Experimental observations have also unambiguously excluded the possibility that the observed QPI pattern are from the Ag(111) substrate states, and proved that they are originating from intrinsic states of silicene. These results pave the way for exploiting anisotropic transport behavior and investigating other exotic physical phenomena of silicene such as quantum spin Hall effect

(QSHE),[3] quantum anomalous Hall effect (QAHE)[24] and superconductivity.[25]

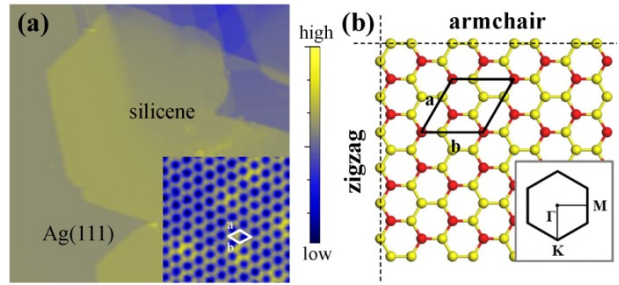

Figure 1. (a) STM image (150×150 nm$^2$, $V_{tip}$ = 1.5 V) of a silicene island crossing Ag(111) steps. The inset shows a high resolution STM image (6×6 nm$^2$, $V_{tip}$ = 1.7 V) of the surface exhibiting a honeycomb structure with lattice constant 0.64 nm. A unit cell is indicated by the white rhombus. (b) Schematic model of silicene-√3 superstructure including zigzag and armchair edges. The red and yellow balls represent silicon atoms with different heights. The black rhombus indicates a √3 unit cell. The inset shows the first Brillouin zone of the 1×1 silicene lattice.

**RESULTS AND DISCUSSION**

Figure 1a is a typical STM topographic image showing a silicene sheet landing across several Ag(111) steps. A zoom-in image of silicene surface is shown in the inset and Figure S1 (see Supporting Information), which exhibits a honeycomb structure with period of 0.64±0.01 nm. This structure corresponds to a √3×√3 superstructure with respect to the 1×1 silicene lattice.[7, 8] The structure of the √3×√3 silicene has been discussed in detail in our previous work [26]. DFT calculations indicated the interaction between √3×√3 silicene and Ag(111) surface is weak van der Waals interaction,[26]. Here we further point out that in our structural model, the number of upper buckled silicon atoms in √3×√3 silicene is less than that of other superstructure of silicene, for example 3×3 silicene. As a consequence, the hybridized sp$^2$ orbitals, and the resulting Dirac fermion characteristics are better preserved in the electronic structures of √3×√3 silicene, which is also proved by our DFT calculations of √3×√3 silicene with Ag substrate (Figure S2, see Supporting Information).

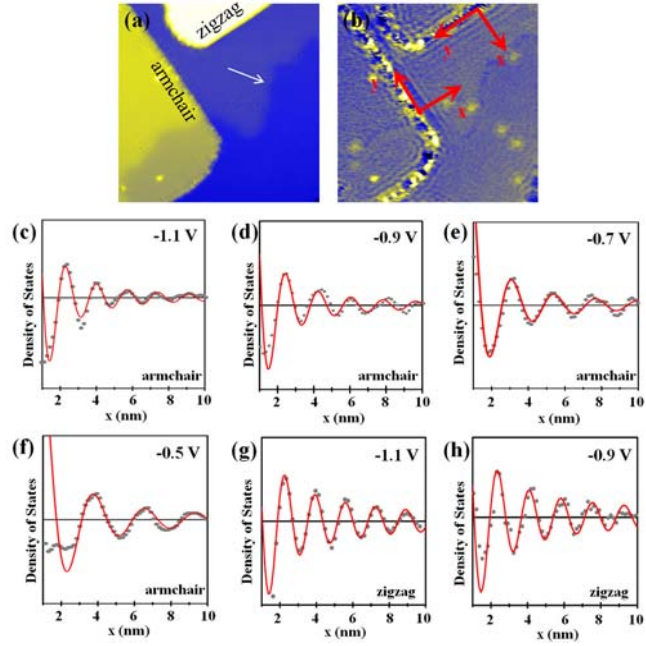

Figure 2. (a) STM image (50×50 nm$^2$, $V_{tip}$ = -1.0 V) including both armchair and zigzag edges of silicene. The white arrow indicate a step edge of Ag(111) beneath the continuous monolayer silicene sheet. (b) dI/dV map ($V_{tip}$ = -1.0 V) at the same area as (a) showing standing wave patterns. We define the direction perpendicular to the step edge as x axis, and that parallel to the step edge as y axis, as illustrated in (b). The position of step edge is set at x=0. (c)-(h) LDOS on monolayer silicene as function of distance from the step edges at different energies. The data were acquired near armchair edge (c)-(f) and zigzag edge (g)-(h) and have been averaged in the y direction to maximize the signal to noise ratio. Grey dots: experimental values; red lines: fits to the data (see text).

Scanning tunneling spectroscopy (STS) records differential tunneling conductance, dI/dV, as a measure of the local density of states at the tunneling energy. On a metallic surface, dI/dV map usually shows Friedel oscillations near step edges or point defects, due to scattering of quasiparticles between the metallic surface states.[27] This has become a powerful tool to probe the surface state characteristics such as symmetry and dispersion. In our experiments, pronounced LDOS oscillation is observed in dI/dV maps near the step edge of a silicene sheet. Figure 2a is an STM topographic image of two silicene islands of 2nd layer silicene on top of 1st layer silicene, where both zigzag and armchair edges are present, perpendicular to each other. The dI/dV map of the same area reveals LDOS oscillations near both types of step edges, as shown in Figure 2b.

In order to understand the origin of the QPI patterns, we have excluded the Ag(111) substrate bands as the origin of the observed standing wave on silicene, as follow. Firstly, although QPI patterns induced by Ag bulk bands had been observed on Ag(110) facet,[28] in the case of Ag(111) QPI patterns induced by bulk bands has never been observed, either on pristine Ag(111) or adsorbate-covered Ag(111). All previously reported QPI patterns on Ag(111) were originated from the Shockley surface state of Ag(111), in the energy range of 0-3 eV above Fermi level.[29] A Shockley surface state is difficult to survive when the Ag(111) is covered with a silicene layer (Figure S3, see Supporting Information). Secondly, silicene on Ag(111) surface exhibit different phases such as 3×3, (√7×√7)R19.1°, and so on.[6,8] If the QPI patterns stem from Ag(111) surface, we should also observe them on other reconstructions. However, we have never observed QPI patterns in any other phases except the √3×√3 silicene.[30] Thirdly, if QPI patterns on silicene originate from the bulk bands of Ag, the wave number should be different for different layer of silicene, because the STM tip at different height from the Ag(111) surface probes bulk bands at different $k_z$. Moreover, when the tip moves farther from the Ag(111), the intensity of standing waves should decreases. In our experiment, both the wave number and the intensity of standing waves are the same for monolayer, bilayer and even trilayer silicene, as shown in Figure S4 (see Supporting Information). Finally, as will be shown later, we observe a strong correspondence between the symmetry of the QPI pattern with the orientation of the silicene domain. Therefore, these long wavelength standing wave patterns are not related to the Ag(111) substrate, but are due to scattering of charge carriers within one silicene Dirac cone, namely intravalley scattering.[31]

It is also notable in Figure 2b that the intensity of the QPI patterns decays as moving away from the step edge. We plot the dI/dV intensity as a function of the distance from the armchair and zigzag step edges at various energies, and some examples are shown in Figure 2c-2h. Here the direction normal to the step edge is defined as *x* and the direction parallel with the step edge is defined as *y*. The dI/dV signal has been averaged in y direction to maximize the signal-to-noise ratio. Clear oscillatory and decay behavior of dI/dV signal along *x* axis are observed. To quantify the decay behavior, the data were fitted using an exponential decaying equation $\delta\rho(x) \propto cos(2kx+\varphi)x^{\alpha}$, which is generally applicable for Friedel oscillations of LDOS in a surface state band,[27] where *k* is the wave vector of the standing wave, $\varphi$ is a phase shift associated with the scattering potential, and $\alpha$ is the decay factor. Other decay channels such as electron-electron interaction, interbands transition have been ignored by treating the lifetime of quasiparticles as infinite (see Supporting Information). The data could be fitted very well by fitting parameter $\alpha$ = -1.5 for armchair edge ($\Gamma$-K direction), and $\alpha$ = -1.0 for zigzag edge ($\Gamma$-M direction).

The LDOS oscillation is a result of quasiparticle scattering (QS) between different points of constant energy contour (CEC). In order to understand our observations, it is helpful to compare our case with three model systems: conventional two dimensional electron gas (2DEG), graphene, and 3D topological insulators (TIs). In conventional 2DEG with parabolic momentum-energy dispersion, the CEC is a perfect circle located at the center of Brillouin zone (BZ). The distribution of QS is thus isotropic, and will show a circle in the momentum space. When scattered from step edges, the decay of QPI patterns as a function of the distance obeys a power law with decay factor of -0.5, *i.e.*, $\delta\rho(x) \propto cos(2kx+\varphi)x^{-0.5}$.[27] In case of graphene, there are six Dirac cones at the K and K' points of BZ. Charge carriers scattered between Dirac cones (intervalley scattering) or within single Dirac cone (intravalley scattering), giving rise to short-wavelength ($\sqrt{3}\times\sqrt{3}$)R30° with respect to 1×1 lattice) or long wavelength interference patterns in real space, respectively.[31] Moreover, for intravalley scattering, one has to take into account the carrier chirality. The projection of pseudospin of quasiparticle in the direction of the momentum defines the chirality.[32] Electrons (or equivalently holes) of opposite direction within one Dirac cone have opposite pseudospins.[21] The standing wave near a step edge is due to scattering between states at the ``stationary points'' of CEC,[33,34] which have anti-parallel pseudospins. This results in suppression of backscattering and a faster decay ($\alpha$ = -1.5).[19] Finally, in the case of 3D TIs such as $Bi_2Se_3$ or $Bi_2Te_3$, only one Dirac cone exists at the BZ center[35,36] and scattering takes place at different points of the CEC. Similarly, due to the real spin texture along the CEC, backscattering is suppressed. Stronger decaying factors, such as -1.5 and -1 had been reported for 3D TIs.[34,37]

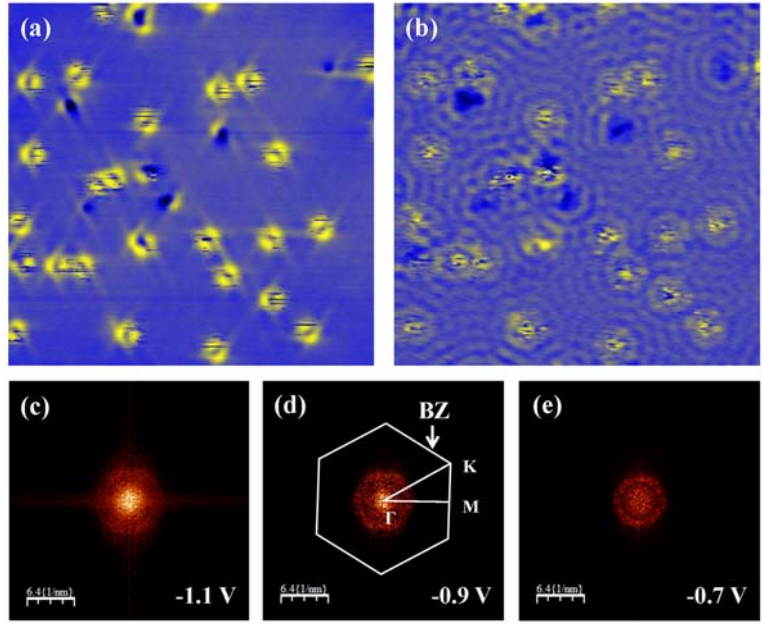

Figure 3. (a) STM image (45×45 nm$^2$, $V_{tip}$ = -2 V) of the hydrogenated silicene surface. (b) dI/dV map ($V_{tip}$ = -0.9 V) at the same area of (a) showing standing wave patterns induced by hydrogenated sites. (c)-(e) k-space maps obtained by Fourier transform of dI/dV maps at different voltages: -1.1 V in (c), -0.9 V in (d) and -0.7 V in (e). A hexagon of 1×1 BZ is superimposed in (d) to show the directions.

The observation of strong decay of LDOS oscillation in silicene suggests that the quasiparticle in silicene is chiral; otherwise the decay of QPI from a step edge should obey a power law with a decay factor of -0.5, as that in conventional 2DEG systems.[27] The faster decay of LDOS oscillation should be due to scattering of electrons between states with non-parallel or even anti-parallel pseudospins. In Bi$_2$Te$_3$, the decay factors are -1.5 in Γ-K direction and -1.0 in Γ-M direction.[20] This high similarity lead us to consider that the Dirac cone of silicene is also hexagonally warped to produce strong and anisotropic decay of QPI patterns.

To validate our hypothesis and further investigate the properties of the Dirac cone of silicene, we turned to QPI from point defects.[38] Since our silicene surface is typically very smooth and defect-free, the QPI patterns are hardly observable on flat terraces due to the lack of scattering centers. To create scattering centers, we deposited 0.001 monolayer (defined as single layer of silicene with lattice constant 0.38 nm) hydrogen atoms on silicene, which are adsorbed on top of Si atoms to form Si-H bonds. The hybridization of Si atoms is thus changed from sp$^2$ to sp$^3$,[39] which will change the local potential and create a scattering center. A typical STM topographic image of silicene after hydrogenation is shown in Figure. 3a. Each eyelike protrusion corresponds to one H atom chemically adsorbed on one Si atom. After hydrogen adsorption the typical dI/dV map now shows significant QPI patterns, as shown in Figure 3b.

A striking aspect of the map is that the waves around point defects exhibit a hexagonal rather than circular shape. This unique feature can be revealed more clearly by fast Fourier transformation (FFT) of dI/dV maps into k space. Figure 3d is the FFT result of Figure 3b, showing a hexagon in the center. Comparing with the surface BZ of 1×1 lattice of silicene, we found the edges of hexagon in Γ-K direction, while the vertices in Γ-M direction. We obtained similar feature in k-space maps in the energy range from 0.4 eV to 1.2 eV above the Fermi energy. Figure 3c and 3e show other two k-space maps taken at -1.1 V and -0.7 V respectively, both of which exhibit hexagonal features.

It is worth to note that the √3×√3 silicene can have different rotational domains, 0°, 30° and ±10° with respect to the Ag(111) substrate, indicating that this phase is incommensurate with the substrate.[26] Importantly, the directions of hexagons are always along the orientation of silicene, not the Ag(111) substrate (Figure 4). If the standing wave we observed on silicene does originate from the bulk bands of Ag, the orientation of the hexagon must be locked to the crystallographic direction of Ag(111). This is another strong evidence to rule out the possibility of QPI patterns originating from Ag bands.

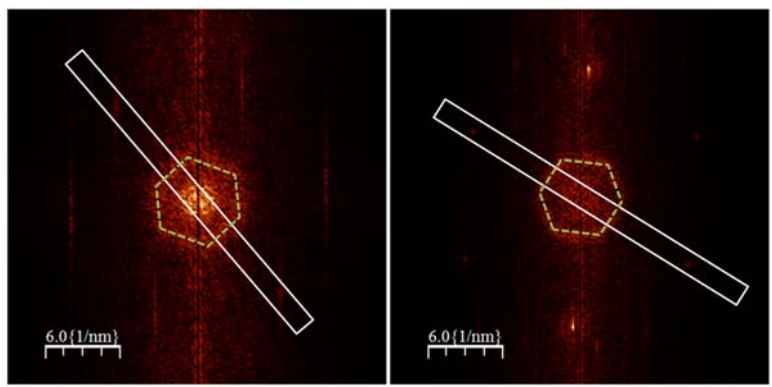

Figure 4. Two FFT images obtained on two silicene islands with different orientations on same Ag single crystal

surface. The dash lines guide eyes to show the hexagonal shapes of distribution of scattering vectors in q-space. The white rectangles mark the orientations of silicene.

In graphene, the CEC of Dirac cone at K and K' points in BZ at low energy is isotropic and circular, but it is trigonally warped at higher energies.[40] The orientation of warping is different for K and K' points. It is thus easy to assume that similar trigonal warping may exist in the Dirac cone of silicene. However the trigonally warped CECs cannot explain the hexagonal shape of LDOS oscillations observed in both real space and k space for silicene. In order to better understand the Dirac cone warping in silicene, we studied the electronic band structure of silicene using first principle calculations. The structural model of our √3×√3 honeycomb silicene on Ag(111) has been understood recently,[26] as shown in Figure 5a. The band structure performed by DFT calculations has also been shown in Ref. 26. It should be noted due to the formation of the (√3×√3)R30° superstructure on silicene surface, the K and K' points in the BZ of (1×1) phase are folded onto the Γ point of the BZ of the (√3×√3)R30° superstructure (shown in Figure 5b). As a consequence, the six Dirac cones with different orientation of warping at different K and K' points of (1×1) phase are folded onto the Γ point of √3×√3 structure and give rise to one Dirac cone, which should become hexagonally warped. The results of DFT calculation shows that linear energy-momentum dispersion is preserved at Γ point with a significant gap stemming from the ultra-buckling Si atoms in √3×√3 phase of silicene.[26] We drew several CECs of Dirac cone at Γ point at energy higher than 0.5 eV above Dirac point, as shown in Figure 5c. It is obviously that the CECs start to hexagonal warping and form hexagons above 0.75eV. The calculation also shows the CEC of Dirac cone is still far away from the boundary of first Brillouin Zone. So the warping effect should not result from the interactions with other bands.

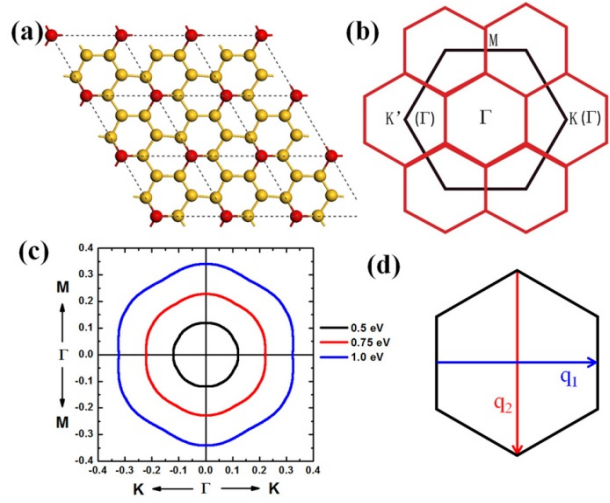

Figure 5. (a) The lattice geometries of (√3×√3)R30° rhombic superstructure of silicene. The red and green balls represent the Si atoms with different heights. (b) The BZ of 1×1 and (√3×√3)R30° silicene are plotted in black and red lines. Note that the Dirac points k and K' in the BZ of 1×1 cell are at the Γ point in BZ of (√3×√3)R30° cell. (c) CECs of Dirac cone of (√3×√3)R30° structures obtained from DFT calculations. The energies are indicated. (d) The schematic illumination of the scattering wave vectors in hexagonal warping CECs of Dirac cone.

Based on the calculated Dirac cone structure, the QPI in silicene can be understood as follow. For a circular CEC, the scattering from states $k_F$ and $-k_F$, namely backscattering, are the most efficient process due to enhanced phase space.[21] This results in a circle with radius $2 k_F$ in the FFT of the dI/dV maps. In our present case, the CEC of Dirac cone is hexagonally warped, resulting in the primary scattering processes occurring between two opposite edges of hexagon ($q_1$ shown in Figure 5(d)). Therefore the scattering wave vector $q$ varies along the opposite edges, and will result in a hexagon with edges toward Γ-K direction and vertices toward Γ-M direction (for BZ of 1×1 structure) in k-space maps (Figure 3c-3e). On the other hand, the backscattering should be suppressed due to the states with antiparallel pseudospins, which will give QPI with faster decay rate. The coincidence of the observed decay factor, -1.5, with the decay factor observed in chiral Dirac fermion systems such as graphene and 3D topological insulators strongly supports our above model.[19,34] Additionally, occurring of the intravalley backscattering requires a pseudospin-flip process induced by strong scattering potentials such as the point defects and step edges. For weaker scattering potential such as substrate step beneath the continuous silicene sheets, (for example the one marked by white arrow in Figure 2a), no QPI patterns had been observed in our experiment, which can be viewed as a result of protection by quasiparticle chirality. Similarly, the scattering near zigzag edge of silicene (along Γ-M direction) should stem from the scattering between opposite vertices of hexagon in CEC ($q_2$ in Figure 5d). The related decay rate -1.0 may be originated from the finite overlaps of pseudospins, which was also observed in the case of topological insulator $Bi_2Te_3$ with a hexagram-shaped CEC of Dirac cone.[34,37] The warped CEC of Dirac cone usually happens at energy far away from the Dirac point. The scattering vectors can be enhanced at the stationary points of the distorted CEC which provide nesting or near nesting condition. That is why we observed the QPI only at energy larger than 0.5 eV above Fermi energy.

**CONCLUSION AND PROSPECTS**

Our present work unambiguously reveals that the Dirac cone in silicene is hexagonally warped, and with chirality. These features make silicene very unique, as compared with its structural analogue, graphene. They have also important indication for exploiting silicene in future electronic devices, because transport properties are dependent on the scattering processes and thus will also be anisotropic. The warping of Dirac cones of silicene provides additional possibility to tune the carrier mobility in silicene based devices.[41,42] Understanding the electronic structure of silicene is helpful to investigate other physical properties, for example superconductivity.[25]

**METHODS**

All our experiments were performed in a home-built low temperature STM with a base pressure of $5\times10^{-11}$ torr. The sample was transferred to the STM chamber immediately after preparation without breaking the vacuum. The clean Ag(111) surface was obtained by repeated cycles of Ar ion sputtering and annealing. Silicene film was prepared by depositing 0.8 monolayer (ML) silicon atoms on the Ag(111) surface at substrate temperature of 500 K. The deposition flux of silicon was kept at 0.05 ML/min. $H_2$ molecules are cracked by a tungsten filament kept at approximately 1700 K and then dosed onto silicene at room temperature. All STM topographic images were measured in constant current mode with an electrochemically etched tungsten tip, and the dI/dV signal was acquired using standard lock-in technique by superimposing a small sinusoidal modulation (20 mV and 676 Hz) to the tip bias voltage. All our data were obtained at 77 K. The details of DFT calculations are same to Ref.[26].

*Acknowledgement:* This work was supported by the MOST of China (Grants No. 2012CB921703, 2013CB921702, 2013CBA01600), and the NSF of China (Grants No. 11174344, 91121003, 11074289).Y. Yao is supported by MOST of China (Grant No. 2011CBA00100) and NSF of China (Grants No. 11174337 and 11225418)

*Supporting Information Available:* The spatial distribution of standing waves on Ag(111) and mutilayers silicene, the DFT calculated electronic structures of √3×√3 silicene on Ag(111) surface, and detailed description of decay of standing wave patterns. This material is available *via* the Internet at http://pubs.acs.org.

TOC

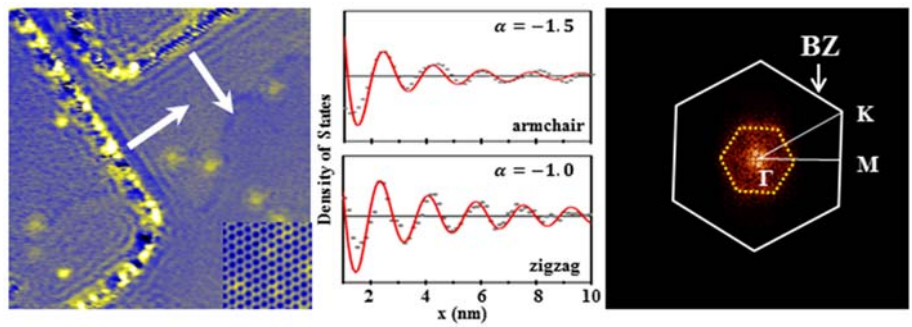